\documentclass[grl]{agu2001}

\usepackage{graphicx}

\authorrunninghead{
YOSHIDA AND KAGEYAMA
}

\titlerunninghead{MANTLE CONVECTION WITH YIN-YANG GRID}

\authoraddr{
Masaki Yoshida and Akira Kageyama,
Earth Simulator Center, Japan Agency for Marine-Earth Science and Technology, 
3173-25 Showa-machi, Kanazawa-ku, Yokohama, Kanagawa 236-0001, Japan.
(myoshida@jamstec.go.jp; kage@jamstec.go.jp)
}
\begin{document}

%
%


\title{
Application of the Yin-Yang grid to a thermal convection of 
a Boussinesq fluid with infinite Prandtl number 
in a three-dimensional spherical shell
}
%

%
%


\author{Masaki Yoshida and Akira Kageyama}

\affil{
Earth Simulator Center, Japan Agency for Marine-Earth Science and Technology, Yokohama, Japan.
}

\begin{abstract}
A new numerical finite difference code has been developed to solve 
a thermal convection of a Boussinesq fluid with infinite 
Prandtl number in a three-dimensional spherical shell. 
A kind of the overset (Chimera) grid named ``Yin-Yang grid'' 
is used for the spatial discretization. 
The grid naturally avoids the pole problems 
which are inevitable in the latitude-longitude grids. 
The code is applied to numerical simulations of mantle convection 
with uniform and variable viscosity. 
The validity of the Yin-Yang grid for the mantle convection simulation is confirmed.
\end{abstract}

%
%

%

\begin{article}

\section{Introduction}
From the middle of 1980s,  
numerical simulation codes for the thermal convection 
with infinite Prandtl number in three-dimensional (3-D) spherical shells  
have been developed to solve the mantle convection of terrestrial planets.  
The discretization methods employed in these codes can be divided into three categories; 
the spectral method  
[{\it Machetel et al.}, 1986; {\it Glatzmaier}, 1988;
{\it Bercovici et al.}, 1989; {\it Zhang and Yuen}, 1995; {\it Harder and Christensen}, 1996],  
the finite element (FE) method  
[{\it Baumgardner}, 1985; {\it Bunge and Baumgardner}, 1995; {\it Zhong et al.}, 2000; 
{\it Tabata and Suzuki}, 2000; {\it Richards et al.}, 2001], 
and the finite volume (FV) method [{\it Ratcliff et al.}, 1996; {\it Iwase}, 1996]. 
The spectral method, which can be an effective method for spherical flows 
[e.g., {\it Fornberg}, 1996; {\it Fornberg and Merrill}, 1997], 
had found to be unsuitable to mantle convection 
simulations because of intense spatial variation of the viscosity of mantle rock. 
A new method based on multilevel wavelet algorithm [{\it Vasilyev et al.}, 1997] 
can treat the spatially localized physical properties and 
has a great potential usefulness in mantle convection simulations. 
Its application to a spherical shell model is, however, 
still remains a challenging task. 
Among the grid-based FE, FV and finite difference (FD) schemes,  
the FV and FD methods are more desirable than FE for massively parallel vector computers 
because of their feasibility of optimization. 
Another advantage of the FD method is its flexibility; 
the extension to higher-order schemes, 
which might be important to obtain accurate solutions  
of thermal convection with very large Rayleigh numbers [e.g., {\it Larsen et al.,} 1997], 
is relatively easy. 

One of the most popular computational grids in the spherical polar coordinates 
($r, \theta, \phi$) is latitude-longitude ($\theta, \phi$)-grid, 
which is defined by intersections of latitude and longitude circles on a sphere (Fig.\ 1a).  
It is widely recognized that the ($\theta, \phi$)-grid has 
the ``pole problems'' that refer to two different kinds of difficulty 
in numerical calculations; 
one is the coordinate singularity on the poles ($\theta = 0, \pi$); 
and the other is the grid convergence near the poles. 
The pole problems have been considered as serious difficulties 
in the community of mantle convection simulation.
To avoid the coordinate singularity, special cares  have to been taken. 
In the FV method, for example, 
all the physical variables are arranged not to reside 
on the pole grids [{\it Ratcliff et al.}, 1996; {\it Iwase}, 1996]. 
The problems of the grid convergence is more serious than 
the coordinate singularity: 
It causes not only the grid redundancy, 
but also the severe restriction on the time-step due to 
the Courant-Friendrichs-Levy (CFL) condition. 
In the ($\theta, \phi$)-coordinates, 
the grid spacing on the spherical surfaces is extremely non-uniform 
as Fig.\ 1a shows. 
The largest grid spacing $\Delta X$ is given in the equator; $\Delta X = 2\pi/N_{\phi}$, 
where $N_{\phi}$ is the grid number in the $\phi$-direction, 
while the smallest grid spacing $\Delta x$ is given at the nearest latitude to the poles; 
$\Delta x = r \sin(\pi/N_{\theta}) \times (2\pi/N_{\phi}) 
\sim 2\pi^2 r / (N_{\theta}N_{\phi})$, where 
$N_{\theta}$ is the grid number in the $\theta$-direction.  
So the ratio $\Delta X/\Delta x \sim N_{\theta}/\pi$ 
increases in proportional to the grid number. 
This means that the time-step restriction becomes extremely severe for 
large scale simulations with fine grids.  
To avoid the impractically small time-step, 
one has to invoke quasi-uniform grid spacing over the sphere. 
The FE based codes referred above employed carefully designed grid cells 
for that purpose.  
For example, a FE mantle convection code 
named CitcomS has nearly uniform resolution 
in both polar and equatorial regions [{\it Zhong et al.}, 2000].  
However, a FD or FV based mantle convection code that overcomes 
both of the pole singularity and the grid convergence have not been reported so far.

Here we employ a new grid system for spherical shell geometry, 
named ``Yin-Yang grid'', which has been proposed recently by {\it Kageyama and Sato} [2004]. 
The Yin-Yang grid is composed of two component grids 
that have exactly the same shape and size (Fig.\ 1b). 
They partially overlap each other on their boundaries (Fig.\ 1c). 
Following the overset (Chimera) grid method 
[{\it Chesshire and Henshaw}, 1990], 
data on the boundaries of the component grids are matched by interpolation.   
A component grid of the Yin-Yang grid is actually a low latitude part 
of the ($\theta, \phi$)-grid. 
As it is apparent in Fig. 1b, 
the Yin-Yang grid has neither a coordinate singularity, nor grid convergence; 
the grid spacings are quasi-uniforms on the sphere 
(see {\it Kageyama and Sato} [2004] for more details on this grid). 

In this paper, we apply the Yin-Yang grid for the numerical simulation of mantle convection.  
To confirm the validity of the Yin-Yang grid, 
we have performed benchmark tests with published numerical codes 
for steady convections. 
We also apply the Ying-Yang grid for time-dependent mantle convections 
with uniform and variable viscosity.

%
%


%
%


\section{Model and Numerical Methods}
We model the mantle convection as a thermal convection 
of a Boussinesq fluid with infinite Prandtl number 
heated from bottom of a spherical shell. 
The ratio of the inner radius ($r=r_0$)
and the outer radius ($r=r_1$) is 0.55. 
The normalization factors for the non-dimensionalization of 
the length, velocity, time 
and temperature are $\hat{r}_1$ = 6371 km (the Earth's radius), 
$\hat{\kappa}/\hat{r}_1$, 
${\hat{r}_1}^2 / \hat{\kappa}$ and 
$\Delta \hat{T} = \hat{T}_{bot} - \hat{T}_{top}$, respectively,  
where 
$\hat{\kappa}$ is the thermal diffusivity, 
and $\hat{T}_{bot}$ and $\hat{T}_{top}$ are the temperatures 
on the bottom and top surfaces. 
The hat stands for dimensional quantity. 
The non-dimensional equations of mass, momentum, and energy conservation 
governing the thermal convection are,  
\begin{equation}
\mathbf{\nabla} \cdot \mathbf{v} = 0,
\end{equation}
\begin{equation}
0 = - \mathbf{\nabla} p + \mathbf{\nabla} \cdot ( \eta \mathbf{\dot{\epsilon}} ) + Ra \zeta T \mathbf{e}_r, 
\end{equation}
\begin{equation}
\partial_t T = \mathbf{\nabla}^2 T  - \mathbf{v} \cdot \mathbf{\nabla} T,
\end{equation}
where $\mathbf{v}$ is the velocity vector,  
$p$ the dynamic pressure, $T$ the temperature, $t$ the time, 
$\mathbf{\dot{\epsilon}}$ the strain-rate tensor, 
and $\mathbf{e}_r$ is the unit vector in the $r$-direction. 
The constant parameter $\zeta$ is $({\hat d}/{\hat r_1})^3 = 0.45^3$, 
where $\hat{d}$ is the thickness of the shell, 2890 km (the Earth's mantle). 
We assume that viscosity $\eta$ depends only on temperature;  
$\eta(T) = \eta_{ref} \exp \left[ -E \left( T - T_{ref} \right) \right]$, 
where $T_{ref}$ is the reference temperature, 
and $\eta_{ref}$ is the reference viscosity at $T_{ref}$. 
The parameter $E$ denotes the degree of viscosity contrast 
between the top and bottom surfaces. 
The viscosity contrast across the spherical shell is defined by 
$\gamma_{\eta} \equiv \eta(T_{top}) / \eta(T_{bot}) = \exp(E)$. 
The Rayleigh number is defined by   
$Ra \equiv \hat{\rho} \hat{g} \hat{\alpha} \Delta \hat{T} \hat{d}^3 / 
\hat{\kappa} \hat{\eta}_{ref}$,
where  
$\hat{\rho}$ is the density, 
$\hat{g}$ the gravitational acceleration, and 
$\hat{\alpha}$ is the thermal expansivity. 
The mechanical boundary conditions at the top  
and bottom surface are immpermiable and stress-free. 
The boundary conditions for $T$ are 
$T_{bot} = 1$ and $T_{top} = 0$. 

We use the collocated grid method 
[e.g., {\it Ferziger and Peri\'{c}}, 2002]; 
all the primitive variables, $\mathbf{v}$, $p$ and $T$, are defined on the same grid points. 
Equations (1)-(3) are solved by the FD discretization with second-order accuracy. 
The SIMPLER algorithm [{\it Patankar}, 1980; {\it Ferziger and Peri\'{c}}, 2002] is applied 
to solve $\mathbf{v}$ and $p$ from eqs.\ (1) and (2). 
The Crank-Nicolson method is used in eq.\ (3) for the time stepping. 
The upwind difference method is applied for the advection term in eq.\ (3). 
With the Yin-Yang grid method, 
we simultaneously solve eqs.\ (1)-(3) for each component grid.  
We use a successive over-relaxation (SOR) method 
as the iterative solver 
required in the SIMPLER algorithm and the energy equation.
The horizontal boundary values of each component grid are determined by 
linear interpolation from the other component grid. 
The interpolation is taken at each SOR iteration.  
(We confirmed that the interpolation procedure has no 
numerical mischief on the calculations.)   
The grid size is $102 \times 102 \times 204$ 
(in $r$-, $\theta$-, and $\phi$-directions). 
We have confirmed that this size is enough to resolve 
all the convections studied in this paper. 
Time development of the convection is calculated 
until averaged quantities, such as Nusselt number and 
root-mean-square velocity, become stationary. 

\section{Benchmark Tests}

The thermal convection in the spherical shell with infinite Prandtl number 
has two stable solutions with polyhedral symmetry when the Rayleigh number is low 
[e.g., {\it Schubert et al.}, 2001]. 
The two solutions are found by linear theory  
[{\it Busse}, 1975; {\it Busse and Riahi}, 1982]
and confirmed by numerical simulations 
[{\it Bercovici et al.}, 1989; {\it Ratcliff et al.}, 1996]: 
One solution is a convection with the tetrahedral symmetry which has four upwellings;   
the other has the cubic symmetry with six upwellings. 
To confirm these symmetric solutions and their stabilities,  
we performed two simulations with different initial conditions of temperature field;   
$T(r, \theta, \phi) = T_{cond}(r) + T_{prtb}(r, \theta, \phi)$, 
where $T_{cond}(r) = r_0 (r_1 -r) / r (r_1 - r_0)$ 
is the purely conductive profile, 
i.e., $\mathbf{\nabla}^2 T_{cond}(r) = 0$, 
with the thermal boundary conditions given above. 
The perturbation term $T_{prtb}(r, \theta, \phi)$ is given by,  
\begin{eqnarray}
T_{prtb}(r, \theta, \phi) =  \lambda \left[ Y_3^{2}(\theta, \phi) + \Omega(\theta, \phi) \right] \sin \pi (r - r_0),
\end{eqnarray}
for the tetrahedral symmetric solution, and 
\begin{eqnarray}
T_{prtb}(r, \theta, \phi) =  \lambda \left[  {Y_4}^{0} (\theta, \phi) 
                           + \frac{5}{7} {Y_4}^{4} (\theta, \phi)  + \Omega(\theta, \phi) \right]  \nonumber \\
                       \times \sin \pi (r - r_0),
\end{eqnarray}
for the cubic symmetric solution, 
where ${Y_\ell}^m (\theta, \phi)$ is the fully normalized spherical harmonic functions  
of degree $\ell$ and order $m$.  
The  ${Y_\ell}^m (\theta, \phi)$ terms in eqs.\ (4) and (5) determine the solution's symmetry. 
The other term $\Omega (\theta, \phi)$ is for secondary perturbation. 
We set $\Omega (\theta, \phi) 
= \omega \sum_{\ell=1}^{12} \sum_{m=0}^{\ell} {Y_{\ell}}^m (\theta, \phi)$. 

We have performed benchmark tests 
with published numerical mantle convection codes 
that employed various numerical schemes. 
Following {\it Richards et al.} [2001] and {\it Ratcliff et al.} [1996], 
we performed simulations of uniform ($\gamma_{\eta}=1$) 
and variable ($\gamma_{\eta}=20$) viscosity convections 
with both the tetrahedral and cubic steady symmetries 
when $\lambda=10^{-1}$ and $\omega=0$ (i.e., no secondary perturbations). 
The Rayleigh number $Ra_{1/2}$ is defined by the reference viscosity 
$\eta_{ref}$ at $T_{ref} = 0.5$ [{\it Ratcliff et al.}, 1996].  
Nusselt number at the surface and root-mean-square velocity of entire domain 
were calculated on convections  
at $Ra_{1/2} = 2.0 \times 10^3 \sim 1.4 \times 10^4$. 
The results of the benchmark tests are summarized in Table 1.  
In spite of the differences of the discretization methods, numerical techniques, 
and number of grid points among the codes, 
we found that the results from our code agree well with them within a few percent or even better  
and confirmed the validity of our code. 

\section{Unsteady Convection Problems}

The steady convections become time-dependent 
when the Rayleigh number is increased. 
Since the Earth's mantle is obviously time-dependent convection 
with high Rayleigh number,  
the transition of convection from steady to unsteady state is important. 
We tried  a series of simulations with various $Ra_{bot}$ 
(the Rayleigh number defined by the reference viscosity at the bottom surface, 
i.e., $T_{ref} = T_{bot}$)
from the critical number 
for convection onset ($\approx 712$) [{\it Ratcliff et al.}, 1996] to $10^5$. 
The perturbation amplitudes $\lambda$ and $\omega$ 
are taken to be $10^{-1}$ and $10^{-3}$, respectively. 
Shown in Fig.\ 2 are the iso-surfaces of temperature at $Ra_{bot}=10^4$ and $10^5$ 
after 200,000 time-steps. 
Figure 2a and 2b indicate that, at $Ra_{bot}=10^4$, 
the convection patterns are in steady states, maintaining each symmetry, 
in spite of the existence of the secondary perturbations in the initial conditions.  
This is consistent with earlier results 
[{\it Bercovici et al.}, 1989; {\it Ratcliff et al.}, 1996] 
in which the secondary perturbation was not explicitly imposed, i.e., $\omega=0$, though.

When $Ra_{bot} = 10^5$, 
the convection patterns become weakly time-dependent. 
The geometrical symmetry in this Rayleigh number is broken.  
This disagrees with the result of {\it Ratcliff et al.} [1996]. 
Notice that, in the right panel of Fig.\ 2b, all the six upwelling plumes 
have the same diameters in our results. 
The corresponding case by {\it Ratcliff et al.} [1996], 
in which a FV scheme on the ($\theta, \phi$)-grid is used, 
shows a symmetric pattern about equator  
and appears to remain in a steady state [cf.\ {\it Ratcliff et al.}, 1996, Fig.\ 6]. 
These observations suggest that the low Rayleigh number convections around $Ra_{bot} = 10^5$ 
are numerically affected by coordinate singularity and the grid convergence  
in the ($\theta, \phi$)-grid. 
On the other hand, the pole effects are removed in our code by making use of the Yin-Yang grid. 

It is known that variable viscosity with strong temperature dependence 
induces drastic effects on the convection pattern  
in 3-D Cartesian model with large aspect ratio 
and also in the spherical shell model 
[{\it Ratcliff et al.}, 1997; {\it Trompert and Hansen}, 1998]. 
To confirm this effect in our model, 
we performed simulations with variable viscosity.   
Taking eq.\ (4) as the initial temperature perturbation,  
we first calculated an isoviscous convection at $Ra_{bot}=10^6$. 
The obtained solution, which is shown in Fig.\ 3a, 
is strongly time-dependent and exhibits complex feature 
in contrast to the case at $Ra_{bot}=10^5$ 
(the right panel of Fig.\ 2a). 
We gradually increased $\gamma_{\eta}$ from $1$ (isoviscous case) up to $10^{3}$. 
We obtained a convection regime that has 
cold and rather thick thermal boundary layer on the top surfaces (Fig.\ 3b).  
The large aspect ratio of convecting cells in this regime is consistent with 
the previous results obtained by the 3-D Cartesian model with large aspect ratio 
as well as spherical shell model 
with moderately strong temperature-dependence of viscosity ($\gamma_{\eta} = 10^3$) 
[{\it Ratcliff et al.}, 1997]. 
Our results show that 
the underlying convection patterns with larger aspect ratio of degree-2 come to dominate.  
The two cells structure that consists of one sheet-like downwelling 
along a great circle of spherical shell  
and two mushroom-shaped upwelling plumes is formed.

\section{Conclusions and Discussion}
We have developed a new numerical simulation code 
to solve the thermal convection of 
a Boussinesq fluid with infinite Prandtl number 
using a second-order FD method on newly devised 
spherical overset grid named Yin-Yang grid. 
The validity of the Yin-Yang grid for the mantle convection simulation is confirmed 
by benchmark tests. 
Our code is powerful and unique FD based code that can solve 
both the uniform and the strongly variable viscosity convections. 
The Yin-Yang grid is suitable to solve the mantle convection problems because
it automatically avoids the pole problems that are inevitable on the ($\theta, \phi$)-grid.  
In the isoviscous case with cubic symmetry at $Ra_{bot}=10^5$, 
the convection pattern has a weak time-dependence in our Yin-Yang grid, 
while it was steady with strange asymmetry of the plume sizes 
between those on the poles and those in the equator 
in the previous FV scheme on the ($\theta, \phi$)-grid. 
This discrepancy might be a consequence of the grid convergence  
near poles in the ($\theta, \phi$)-grid. 
Our result implies that large-scale (low degree) convective structures  
are easily affected numerically by the poles
when ($\theta, \phi$)-grid is employed.  
The quadrulpole convection patterns  
is obtained when large viscosity contrast 
with three orders of magnitude is introduced when $Ra_{bot}=10^6$. 

To follow mantle convection for geophysical time-scale ($\sim$$10^{8}$ years), 
the computational time-step $\Delta t$ is critically important in numerical simulations. 
As we described in section 1, 
the time-step is determined by the CFL condition 
by the smallest grid spacing. 
For ($\theta, \phi$)-grid, $\Delta x (= \Delta x_{\theta \phi})$ is determined by 
the azimuthal grid spacing at the nearest grids to the pole. 
On the other hand for the Yin-Yang grid, $\Delta x (= \Delta x_{YY})$ 
is determined by the azimuthal grid spacing at $\theta = \pi/4$ (or $3\pi/4$). 
Therefore the ratio of time-steps $\Gamma_t$ between two grids is, 
\begin{equation}
\Gamma_t 
\propto \Delta x_{\theta \phi}/ \Delta x_{YY} 
= \sin (\pi/N_{\theta}) / \sin (\pi/4)  
\approx 1.4 \pi / N_{\theta}. 
\end{equation}
Taking $N_{\theta} =$ 102 as employed in this paper, 
$\Gamma_t \approx 0.04$.  
This means that the total computational time is 
significantly reduced by the factor of 1/25 
by making use of the Yin-Yang grid.


%
%

\begin{acknowledgments}
The authors are grateful to Prof. David A. Yuen and an anonymous reviewer 
for their careful reviews and valuable comments. 
All the simulations were performed by Earth Simulator, 
Japan Agency for Marine-Earth Science and Technology. 
\end{acknowledgments}

%
%

\end{article}

\clearpage

%
%
%


\begin{figure}
 \begin{center}
  \includegraphics[width=0.99\textwidth]{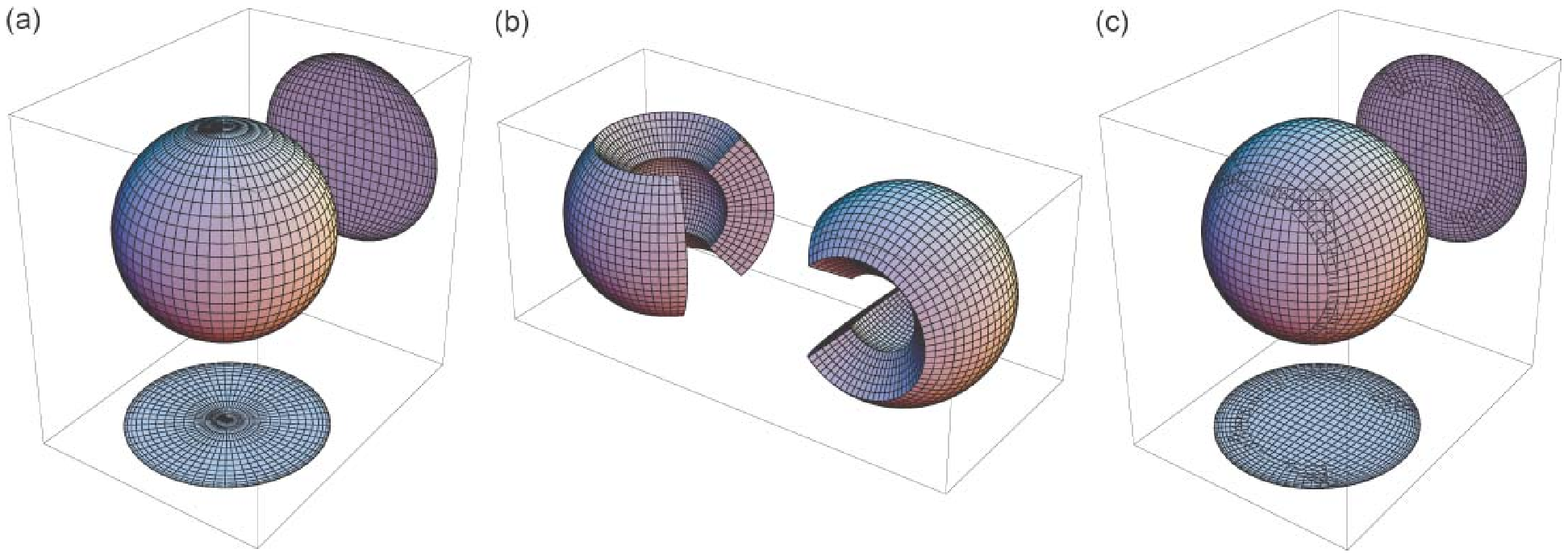}
 \end{center}
 \caption{
The latitude-longitude ($\theta, \phi$)-grid 
and new spherical overset grid named ``Yin-Yang grid''. 
(a) The ($\theta, \phi$)-grid. 
(b) Two component grids of the Yin-Yang grid. 
They are identical (same shape and size); the low latitude part 
($\pi/4 \leq \theta \leq 3\pi/4$, $-3\pi/4 \leq \phi \leq 3\pi/4$) 
of the ($\theta, \phi$)-grid.  
(c) They partially overlap each other at their interface to cover a spherical surface in pair 
(see text and {\it Kageyama and Sato} [2004] for details).
}
\end{figure}

\newpage
\begin{figure}
 \begin{center}
  \includegraphics[width=0.5\textwidth]{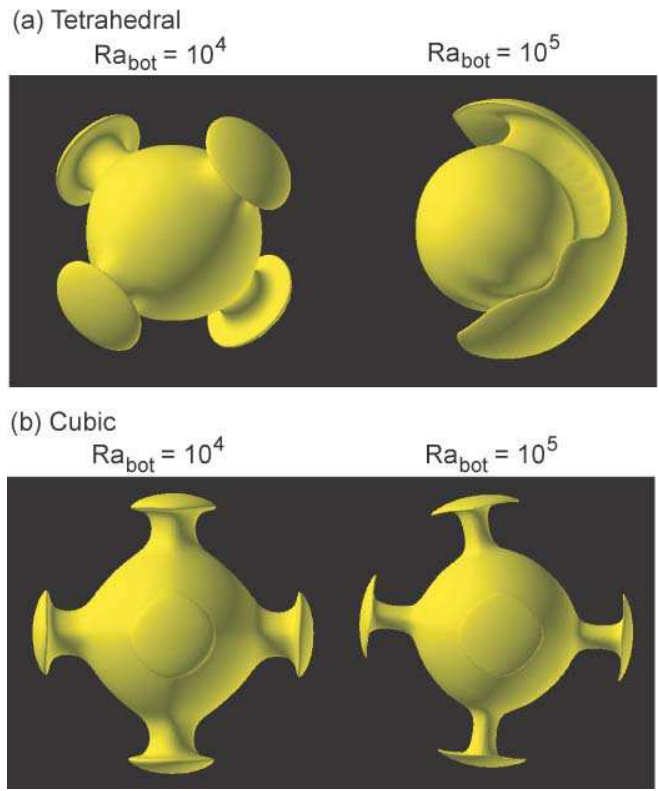}
 \end{center}
 \caption
{
The iso-surface renderings of temperature ($T=0.4$) 
started from the initial conditions of (a) tetrahedral, and (b) cubic symmetries.   
The left and right panels on each figure  
show the cases at $Ra_{bot} = 10^4$ and $10^5$, respectively. 
}
\end{figure}

\newpage
\begin{figure}
 \begin{center}
  \includegraphics[width=0.5\textwidth]{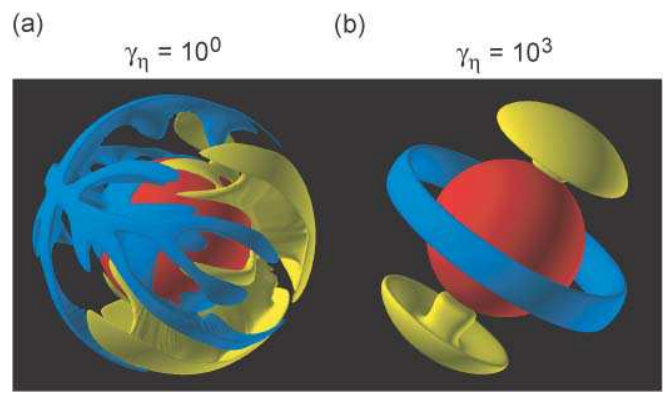}
 \end{center}
 \caption
{
The iso-surface renderings of residual temperature $\delta T$ 
(i.e., the deviation from horizontally averaged temperature at each depth) 
for the cases at $\gamma_{\eta} =$ (a) $10^0$, and (b) $10^3$. 
Blue iso-surfaces stand for $\delta T$ of (a) $-0.10$, and (b) $-0.25$. 
Yellow iso-surfaces for $\delta T$  of (a) $+0.10$, and (b) $+0.25$. 
Red spheres indicate the bottom surface of spherical shell. 
}
\end{figure}

%
%
\clearpage

\begin{table}
\def\ph{\phantom{age }}
\def\pph{\phantom{th}}
\catcode`?=\active \def?{\phantom{0}}
\newbox\dothis
\setbox\dothis=\vbox to0pt{\vskip-1pt\hsize=11.5pc\centering
Error\vss} \caption
{
The benchmark test of Nusselt numbers at the top surface ($Nu$) 
and RMS velocity ($V_{rms}$) of the entire mantle\tablenotemark{a}
}  

\begin{tabular*}
{\textwidth}
{ccc|ccccccccc|cccc}
\hline 
\multicolumn{3}{c}
{
}
&\multicolumn{9}{c}
{
\vrule height 12pt width 0pt 
$Nu$
}
&\multicolumn{4}{c}
{
$V_{rms}$
}
\cr
\cline{4-12}\cline{13-16}
\cr
T/C & 
$Ra_{1/2}$ & $\gamma_{\eta}$ & \shortstack{Gl88\\(SP)} & \shortstack{Br89\\(SP)} & \shortstack{HC96\\(SP)} & 
                            \shortstack{Rt96\\(FV)} & \shortstack{Iw96\\(FV)} & \shortstack{Zh00\\(FE)} &
                            \shortstack{TS00\\(FE)} & \shortstack{Rc01\\(FE)} & \shortstack{YK04\\(FD)} &
\shortstack{Rt96\\(FV)} & \shortstack{Iw96\\(FV)} & \shortstack{TS00\\(FE)} & \shortstack{YK04\\(FD)} \\ 
\hline
T & 2.0e3 &?1& -      & 2.2507 & -      & 2.1740 & 2.18?? & 2.218? & 2.2432 & -      & 2.2025? & 12.14?? & 12.4710 & 12.5739 & 12.1246 \\ 
T & 7.0e3 &?1& -      & 3.4657 & 3.4957 & 3.4423 & 3.45?? & 3.519? & 3.6565 & 3.4160 & 3.4430? & 32.19?? & 32.4173 & 32.9360 & 32.0481 \\
T & 1.4e4 &?1& 4.2820 & -      & 4.2818 & 4.2028 & -      & -      & -      & 4.2250 & 4.2395? & 50.27?? & -       & -       & 50.0048 \\
T & 7.0e3 &20& -      & -      & -      & 3.1615 & -      & -      & -      & -      & 3.1330? & 25.69?? & -       & -       & 26.1064 \\
C & 3.5e3 &?1& -      & 2.7954 & -      & 2.8306 & 2.80?? & -      & -      & -      & 2.8830? & 18.86?? & -       & -       & 18.4801 \\
C & 7.0e3 &?1& -      & -      & -      & 3.5806 & 3.54?? & -      & -      & -      & 3.5554? & 30.87?? & -       & -       & 30.5197 \\
C & 1.4e4 &?1& -      & -      & -      & 4.4449 & -      & -      & -      & -      & 4.4231? & 48.75?? & -       & -       & 48.1082 \\
C & 7.0e3 &20& -      & -      & -      & 3.3663 & -      & -      & -      & -      & 3.3280? & 25.17?? & -       & -       & 25.3856 \\
\hline
\end{tabular*}
\tablenotetext{a}{
``T/C" denotes the tetrahedral (``T") or cubic (``C") symmetric solutions. 
The abbreviated code names ``Gl88" is for 
{\it Glatzmaier} [1988], ``Br89" {\it Bercovici et al.} [1989],
``HC96" {\it Harder and Christensen} [1996], ``Rt96" {\it Ratcliff et al.} [1996],
``Iw96" {\it Iwase} [1996], ``Zh00" {\it Zhong et al.} [2000], 
``TS00" {\it Tabata and Suzuki} [2000],  
``Rc01" {\it Richards et al.} [2001], and ``YK04" is for our code.
The ``SP" in parentheses under each code name denotes spectral method, 
and see text for ``FV", ``FE" and ``FD". 
(Note that, in this benchmark test, 
the normalization factor used to non-dimensionalize the length is 
the Earth's radius $\hat{r}_1$, not $\hat{d}$.)  
}
\end{table}

\end{document}